\documentclass[sigconf]{acmart}
\usepackage[most]{tcolorbox} 
\usepackage{xcolor}
\usepackage{booktabs}
\usepackage{balance}
\usepackage{multibib}

\newcites{References}{References}
\newcites{Primary}{Primary sources}
\setcopyright{cc}
\copyrightyear{2026}
\acmYear{2026}
\acmConference[EASE 2026]{The 30th International Conference on Evaluation and Assessment in Software Engineering}{9–12 June, 2026}{Glasgow, Scotland, United Kingdom}

\newcommand{\RqOne}{\emph{\texorpdfstring{RQ$_1$}: Where and when was the study published?}}
\newcommand{\RqTwo}{\emph{\texorpdfstring{RQ$_2$}: What were the goals of the study? }}
\newcommand{\RqThree}{\emph{\texorpdfstring{RQ$_3$}: What software artifacts were used in the study?}}
\newcommand{\RqFour}{\emph{\texorpdfstring{RQ$_4$}: What research methodology was used in the study?}}
\newcommand{\RqFive}{\emph{\texorpdfstring{RQ$_5$}: Which human stakeholders' perspectives were considered in the study?}}
\newcommand{\RqSix}{\emph{\texorpdfstring{RQ$_6$}: What were the key findings of the study, about how commit messages influence corrective software maintenance?}}

\begin{document}

\title{On the Use of Commit Messages for Corrective Software Maintenance: A Systematic Mapping Study}

\author{Syful Islam}
\email{syful.islam@telecom-paris.fr}
\orcid{0000-0002-7441-6987} 
\affiliation{\institution{LTCI, Télécom Paris, Institut Polytechnique de Paris}
  \city{Palaiseau}
  \country{France}
}

\author{Stefano Zacchiroli}
\email{stefano.zacchiroli@telecom-paris.fr}
\orcid{0000-0002-4576-136X} 
\affiliation{\institution{LTCI, Télécom Paris, Institut Polytechnique de Paris}
  \city{Palaiseau}
  \country{France}
}

\begin{abstract}
Corrective maintenance is crucial to ensure the quality of software, thereby improving reliability and user experience. In a version control system (VCS), developers write commit messages to document their changes and support later maintenance. Therefore, the utilization of commit messages to accomplish corrective maintenance has become a common practice among software engineering practitioners and researchers. Still, to this day, no secondary study has mapped the research landscape of how commit messages have been used in corrective software maintenance.

We present a systematic mapping study of 97 primary sources published between 2004 and May 2025, where we examine the goals, potential utilization of source code artifacts along with commit messages, methodologies, stakeholders, and the key findings about their influence on corrective maintenance.

Our analysis reveals a growing interest in the usage of commit messages to perform corrective maintenance tasks, in particular for bug analysis and bug fix identification goals. Surprisingly few studies address other themes such as automated program repair, security development practices, etc. We find that the software artifacts most used in combination with commit messages are commit "diffs" and that repository mining, together with natural language processing (NLP) and artificial intelligence/machine learning (AI/ML) are the methodological foundations of studies in this field. Among stakeholders considered in previous studies, developers play the most important role in shaping corrective maintenance practices.

Key findings in previous studies about commit messages establish their significant role in corrective maintenance, due to the fact that they carry crucial information helpful for stakeholders to understand and improve the code base through the software evolution process. Often, though, commit messages lack important information and are not enough to convey the intent of code changes to future readers. Therefore, developers should be aware of commit message contextual richness while committing code changes in VCS.
\end{abstract}




\maketitle

\section{Introduction}
Software maintenance encompasses a broad set of activities (such as implementing new features, fixing bugs, enhancements, etc.) performed on software products to ensure their quality and long-term sustainability in the market~\citeReferences{martin1983software}. According to Swanson~\citeReferences{swanson1976dimensions}, these activities are classified into three categories, namely corrective, adaptive and perfective. Among them, corrective maintenance deals with modification of a software product performed after delivery to correct discovered bugs in source code~\citeReferences{swanson1976dimensions, canfora2001software}. Corrective maintenance derives its importance not only from the huge costs it induces but also in the way it affects customer relations and ultimately, the revenue of the software company~\citeReferences{alaranta2012knowledge, butt2022software, morovati2024bug}. 

During corrective maintenance, developers need to understand the code changes enough to analyse problems, identify the bugs and determine how they should be fixed without breaking anything~\citeReferences{vans1999program}.
In VCS, commit messages play a crucial role in facilitating collaborative software development and might be the only source of information left for future developers to understand what
changes were made and why~\citeReferences{xue2024automated, li2023commit, tian2022makes}. Therefore, the usage of commit messages to accomplish corrective maintenance has become a common practice within the software engineering community.

To improve corrective software maintenance, researchers investigate the feasibility of commit message usage in combination with various artifacts such as code changes and apply different techniques, namely repository mining, NLP, AI/ML, and manual inspection. These techniques are used in various areas such as bug analysis, prediction, program repair, patch identification, etc. However, the landscape of commit message usage in corrective maintenance remains fragmented, with various goals and methodologies being employed.  Researchers have conducted numerous empirical studies and propose various tools, yet little work has been done to comprehensively map this field and synthesize the findings about how commit messages influence corrective maintenance. To fill this knowledge gap, we conducted and present in this work a systematic mapping study that provides an overview of historical trends, assesses the role of commit messages in corrective maintenance, and identify areas where further research is required.

\textit{Paper structure.} Section~\ref{sec:methodology} details the methodology of our study. We focus on fundamental research questions, such as where and when the research was published, and the goals that drive researchers to engage in corrective software maintenance using commit messages. In addition, we examine the potential utilization of diverse source code artifacts along with commit messages, their usage patterns, methodology, stakeholder perspectives, and the key findings about how commit messages influence corrective maintenance activities.
We followed a systematic methodology with defined inclusion/exclusion criteria, screening steps, backward snowballing, validated and refined via discussion between authors.
This led to the selection of 97 primary studies, which were then categorized based on various dimensions related to our research questions. Section~\ref{sec:results} presents our main results, which can be summarized as:
\begin{itemize}
    \item There is consistent and growing research interest in the usage of commit messages to address corrective maintenance tasks, with a spike in the 2022--2023 years.
    \item The major goals of corrective maintenance pursued by mining commit messages are bug analysis and bug fix identification; surprisingly few studies address others such as automated program repair, security development practices, etc.
    
    \item The most dominant recurring source code artifact usage pattern is $\{$commit messages, code changes$\}$ within the software engineering community where repository mining serves as the methodological foundation, often supported by 
    $\{$NLP, AI/ML$\}$ and occasionally complemented by expert manual inspection.

    \item Among human stakeholders considered in previous work, developers appeared most frequently in our primary sources, indicating their central role and influence in the corrective software maintenance strategies.
    
    \item Our findings reveal that commit messages play a significant role in corrective maintenance activities since they carry crucial information helpful for stakeholders (i.e., developer, maintainer, and researcher) to understand and improve code base through software evolution process. However, commit messages often lack necessary information and are not always sufficient to represent the intent of code changes.
\end{itemize}

In Section~\ref{sec:implication}, we present a set of key recommendations for stakeholders such as developers, maintainers, and researchers supported by our primary sources. Before concluding (Section~\ref{sec:conclusion}), we discuss threats to the validity of our study in Section~\ref{sec:threats}.

\textit{Contributions and Data Availability.}  This paper presents (i) the first secondary study on the usage of commit messages for corrective software maintenance activities, and (ii) a publicly available replication package~\citeReferences{Replication2026}
containing the dataset of 97 analyzed papers and a structured taxonomy categorizing their key characteristics. All code and data used in this study are included in the package.

\section{Methodology}
\label{sec:methodology}
To conduct this study, we adopt the methodology proposed
by~\citetReferences{petersen2008systematic, petersen2015guidelines} to ensure a systematic and reproducible approach for selecting and analyzing relevant articles.
The methodology is organized around three main phases: developing research questions (Section~\ref{researchquestion}), selection of primary sources (Section~\ref{primanrysourceselection}), and data collection (Section~\ref{datacollection}). 

\subsection{Research questions}
\label{researchquestion} 
To systematically assess the role of commit messages in corrective software maintenance, according to scientific works published so far, we stated the research questions (RQ) listed below, together with their rationales:
\begin{itemize}
    \item \RqOne \\Rationale: grasping the evolution over time and macro-fields of studies that rely on commit messages.
    \item \RqTwo \\Rationale: mapping the ultimate, often unstated, reasons that drive researchers to engage in corrective software maintenance using commit messages.
    \item \RqThree \\Rationale: grasping the potential utilization of diverse source code artifacts in corrective software maintenance, along with commit messages.
    \item \RqFour \\Rationale: understanding which methodological techniques are needed to benefit from commit messages for corrective maintenance needs.
    \item \RqFive \\Rationale: understanding which stakeholders are involved is essential to locate who can later \emph{improve} corrective software maintenance practices. Answer to this RQ will also help in identifying potential biases and gaps in empirical evidence, as well as opportunities to improve collaboration and knowledge transfer.
    \item \RqSix \\Rationale: understanding the key findings from studies provides overview of how commit messages influence corrective maintenance, and also best practices for practitioners.
\end{itemize}

\subsection{Selection of primary sources}
\label{primanrysourceselection}
\begin{figure}[t]
  \centering
  \includegraphics[width=.85\linewidth]{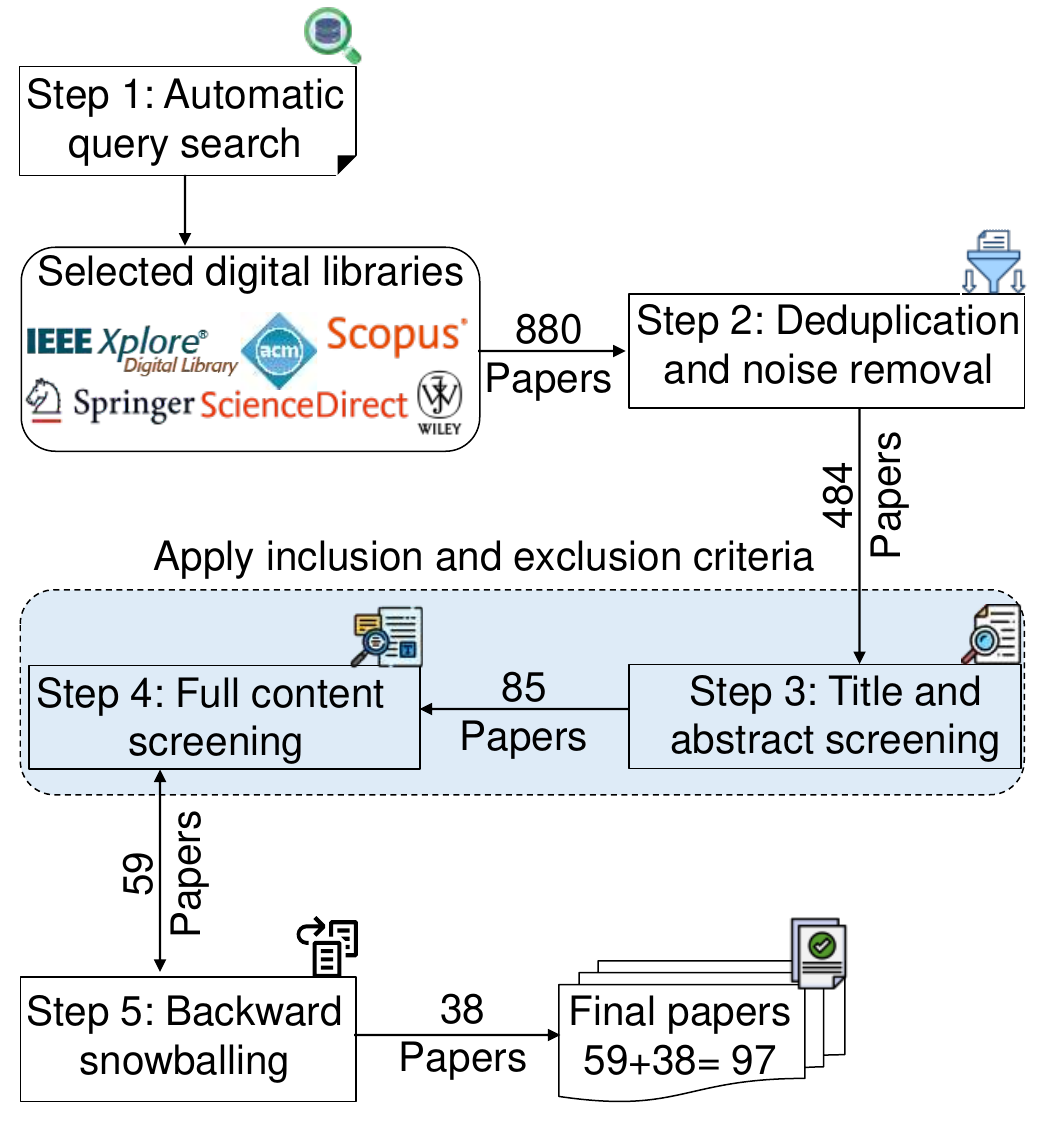}
  \caption{Protocol used for the selection of primary sources.}
  \label{fig:primarysource}
\end{figure}

Figure~\ref{fig:primarysource} illustrates the protocol applied for selecting our primary
sources. Initially, we formulated automatic query search criteria to be used on selected popular online
libraries and executed them. After deduplication and noise removal of obtained results, we applied inclusion and exclusion criteria, titles and
abstracts screening, reviewing full content of the sources, and conducted backward snowballing phase to obtain the final set of primary sources.

\begin{figure}[h!]
    \centering
    \footnotesize
    \begin{tcolorbox}[
        colback=gray!5,     % light background color
        colframe=black,     % border color
        boxrule=0.8pt,      % border thickness
        arc=3pt,            % rounded corners
        left=6pt, right=6pt, top=6pt, bottom=6pt, % padding
        width=\linewidth,
        breakable           % allows content to break across pages
    ]
\begin{verbatim}
{Title:((commit OR commits) AND (message OR messages)) 
OR 
Abstract:((commit OR commits) AND (message OR messages))}
AND 
Abstract:(<domain context terms>)
\end{verbatim}
    \end{tcolorbox}
    \caption{Query template applied on the selected digital libraries. This query searches for the singular or plural form of ``commit'' and ``message'' words in either title or abstract and any domain context terms (i.e., ``software'', ``repository'', ``project'', ``application'', ``code'', ``github'') in abstract to keep the search highly relevant.
}
    \label{fig:query}
\end{figure}

\subsubsection{Step 1: Automatic query search} To identify the key terms for our search query, we adopted the PICO framework~\citeReferences{kitchenham2009systematic}, commonly used in
systematic reviews, which stands for “Population, Intervention,
Comparison, and Outcome”. This approach formulates research
questions and structure as outlined by~\citetReferences{petersen2008systematic}. Our
systematic mapping study focused on the Population and Intervention components. The Population component in our study refers
to software domain context entities. We defined this using the terms: software, repository, project, application, code, github. The Intervention component pertains to the classification processes applied to these entities. We defined this using singular and plural forms of the terms: commit and message. 

We did not include the comparison component because our
study does not involve comparing different methods or interventions against each other. Our objective is to map the existing research on corrective software maintenance as a whole rather than to evaluate or contrast specific approaches. Similarly, we also excluded the outcome component because we are not focusing on specific measurable outcomes from commit message usage in corrective maintenance activities such as accuracy metrics.
In summary, our aim is to identify and categorize the following characteristics of existing studies: data sources, methodologies, and key findings about commit message usage in corrective maintenance activities.

Figure~\ref{fig:query}, presents the search query template, adapted to the syntax and semantics of selected bibliographic database. This query was
applied exclusively to the title and abstract of research articles. We also added domain context terms in query because ``commit or commits'' and  ``message or messages'' keywords are very common in the human psychology literature such as ``No one commits suicide: Textual analysis of ideological practices''; searching for them in article title and abstract returned large amounts of non-relevant articles during our iterations on the search query. Thus, our approach ensures high relevance of the retrieved articles without compromising the breadth of the search. 

We conducted automatic query search on several popular online digital libraries i.e., IEEE, ACM, Scopus, Springer, ScienceDirect, and Wiley similar to a previous study by~\citetReferences{balla2025automatic}. These selected digital libraries represent the
major repositories available in the field of software engineering. For ScienceDirect database, we applied query to the title, abstract, and keywords, as the database natively supports built-in filtering across these fields. In addition, given that Springer database does not support complex queries, we adjusted our search strategy using phrases (i.e.,``commit message'' OR ``commit messages''  OR ``commits'') on article title only to maintain the core focus of the query. To reduce non-relevant articles from Springer database, we compared different search phrase combinations, evaluating the noise ratio in the returned results. Furthermore, while executing query, the whole database was searched instead of specifying time frame aiming to build a more comprehensive dataset on the target topic for systematic mapping study. Thus, we obtained a total of 880 papers from initial automatic query on the selected digital libraries.
\begin{table}[t]
\centering
    \footnotesize
    \caption{{Inclusion and exclusion criteria.}}
    \label{tab:inexcriteria}
\begin{tabular}{@{}p{1.5cm} p{3cm} p{3cm}@{}}
\toprule
\textbf{Parameters} & \textbf{Inclusion} & \textbf{Exclusion} \\
\midrule
Contribution content & Articles that includes contributions related to corrective maintenance & Articles that do not include this feature in the contribution \\
Language & Article written in English & Article not written in English \\
Source type & Must be peer reviewed & Not peer reviewed such as tutorials, editorial notes, abstract, technical reports, thesis, book chapter etc. \\
Accessibility  & Article available in full text & Article not available in full text \\
Extension & If multiple publications of
the same study exist presenting
the same analysis, the latest
version will be included & Articles for which a newer
or more complete version exists \\
\bottomrule
\end{tabular}
\end{table}
\subsubsection{Step 2: De-duplication
and noise removal} 
In this step, our goal is to ensure that the collected papers are both unique and highly relevant to mapping study. To achieve this, we first remove duplicates based on paper titles and DOIs. We then manually inspect the remaining studies to catch any duplicates that may persist due to noise in the article metadata. This filtering ensures that every study we include offers a distinct contribution to the knowledge base. Through this process, we have identified 484 candidate papers for title and abstract screening.

\subsubsection{Step 3: Title and abstract screening} In this step,  we performed screening of the retrieved papers based on title and abstract, as recommended by~\citetReferences{petersen2015guidelines}. This step was performed by the first author, who read
titles and abstracts, and applied inclusion/exclusion criteria listed in Table~\ref{tab:inexcriteria}. Out of the 484 total papers retained thus far, 399 papers were rejected, leaving 85 potentially-relevant papers for full content screening.

\subsubsection{Step 4: Full content screening} In this step, we applied the inclusion/exclusion criteria as listed in Table~\ref{tab:inexcriteria} to the full content of the papers obtained from step 3. The task was collaboratively performed by two authors (i.e., 1st and 2nd author). This task began by discussing and aligning on systematic approach to identify the contribution content and ultimately agreeing to assess whether commit messages were mined as part of the studied methodology and were aligned with corrective maintenance. We did not measure Kappa agreement because the task was conducted collaboratively through continuous discussion and consensus-building, making statistical inter-rater reliability measurement unnecessary~\citeReferences{o2020intercoder}.

Based on the assessment conditions, the first author manually annotated all 85 papers obtained from step 3. The annotation process allowed the first author to mark papers as ``may be'' in case of uncertainty about their inclusion, prompting
a second evaluation by the second author. After mitigating the uncertain cases, we accepted 59 papers relevant to our target study.

\subsubsection{Step 5: Backward snowballing}
To improve retrieval of relevant studies missed in the previous steps,  we incorporated backward snowballing as an additional step. This process was applied to the 59 papers accepted during the selection of primary sources in step 4.
The snowballing phase followed established methodology
used for evaluating primary sources: we began with 59 direct
studies and manually checked the references of each study iteratively. These were processed through three sequential steps: (i) removal of duplicate papers, (ii) document filtering through ``commit message'' keyword searching and (iii) full-content screening. During full-content screening, we found that some studies incorporated commit messages as part of their methodology but did not highlight this in the title or abstract, making it necessary to include these studies in the primary sources. After three successful backward snowballing iterations, we were not able to identify any new papers, indicating that the process had fully converged. Thus, we identified 38 additional studies, bringing the total number of primary sources to 97.

\begin{table}[t]
\centering
\footnotesize
\caption{{Data collection scheme.}}
    \label{tab:datascheme}
\begin{tabular}{@{}p{1.5 cm} p{3.5cm} p{2.5cm}@{}}
\toprule
\textbf{Research Question} & \textbf{Dimension} & \textbf{Scale} \\
\midrule
RQ1 & Year & Time-based scale\\\hline
RQ1 & Venue & Single-choice scale\\\hline
RQ2 & Goal & Single-valued open scale \\\hline
RQ3 & Type of source code artifacts used with commit message  & Multi-valued open scale \\\hline
RQ4 & Type of research methodologies  & Multi-valued open scale \\\hline
RQ5 & Type of stakeholders  & Multi-valued closed scale \\\hline
RQ6 & Key findings about commit message  & Multi-valued open scale \\
\bottomrule
\end{tabular}
\end{table}

\begin{figure}[t!]
  \centering
  \includegraphics[width=1\columnwidth]{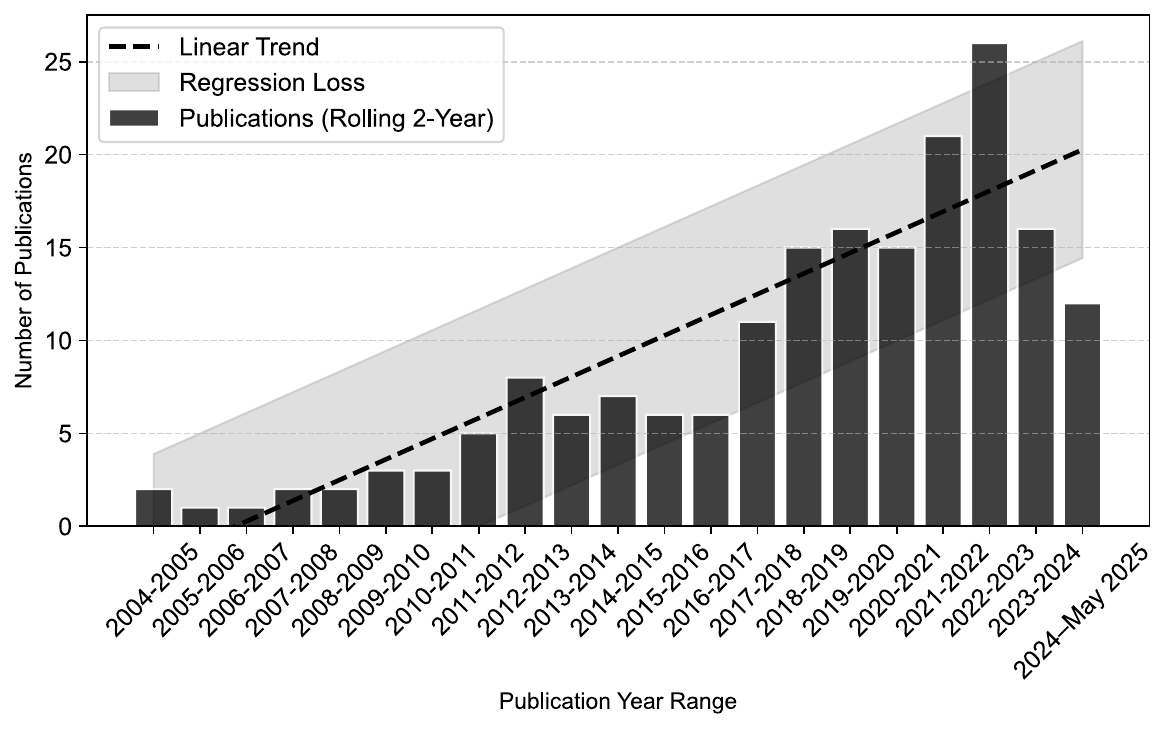}
  \caption{Year-wise publication (rolling 2-year) trend.}
  \label{fig:publicationtrend}
\end{figure}

\subsection{Data collection}
\label{datacollection}
The process of data collection aims to obtain thorough information from
the primary studies corresponding
to the scales, as outlined in Table~\ref{tab:datascheme}. The data collection protocol follows standard rules and practices
commonly employed in software engineering literature reviews~\citeReferences{kitchenham2009systematic, zimmermann2016card}. This involves a series of steps conducted by both authors in three collaborative iterations.

During initial iteration, both authors reviewed the same 25 sample papers from the 97 primary sources.
They reviewed the full text of the papers and extracted the information according to predefined research question dimensions and conducted an initial classification.
Some challenges arise where certain data items are missing, poorly defined, or require inference from context. Such instances are flagged for further investigation in subsequent iterations. 

In the second iteration, authors cross-reference and merge their individual data summaries into an integrated table. Unclear data items are revisited and resolved through discussions and further
research together. For instance, a category labeled ``check together'' is added for the evaluation procedure due to a lack of justification in some studies, requiring additional research to determine the subject system domains and ensure that their interpretations of the different dimensions and scales were aligned. 

In the third iteration, the first author proceeded to read and extract the information from the remaining papers at a flexible yet systematic pace to maintain methodological rigor and held weekly meetings with second author to resolve any arising issues.
Once the information was extracted from all papers, the evaluators met again to homogenize the vocabulary used for the open scales following an open card sorting process~\citeReferences{zimmermann2016card}.
For each open scale, the evaluators listed and consolidated the information extracted from the papers. For instance, for the scale “goal of research”, we consolidated the extracted terms “bug prediction”, “just-in-time defect prediction and localization” under the common term “bug identification and localization”.
Once the exhaustive list of values for each scale was established, the first author applied them to all papers.
For certain open scales, we further analyzed recurring patterns
in the data, grouping them into distinct themes to provide additional depth and structure to the analysis. For instance, for the
scale “Type of source code artifacts used”, we grouped all values related to source code artifacts in version control system (e.g., code changes, issue tracking, vulnerability information from security advisory, etc).

Finally, the results are summarized into integrated table statistics, count the number of studies for each technique associated with data items, and group similar techniques into broader categories.  Both authors participate in comprehensive discussions to finalize data classifications for the study. These classifications and patterns are further discussed in Section~\ref{sec:results}, providing a comprehensive interpretation of the findings.

\section{Results}
\label{sec:results}
In this section, we review and analyze the information extracted from 97 primary sources following the scales and dimensions identified for each research question in Table~\ref{tab:datascheme}.
For each dimension, the number of studies matching it is indicated in parentheses.
The raw data, analysis scripts, and plots discussed in this section are available from the reproducibility package~\citeReferences{Replication2026}.
\begin{figure}[t]
  \centering
  \includegraphics[width=0.8\columnwidth]{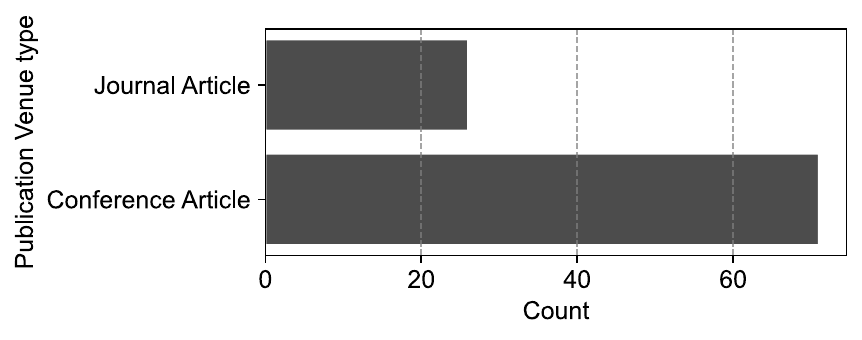}
  \caption{Publication venue types.}
  \label{fig:publicationvenuestypes}
\end{figure}

\begin{figure}[t]
  \centering
  \includegraphics[width=0.80\columnwidth]{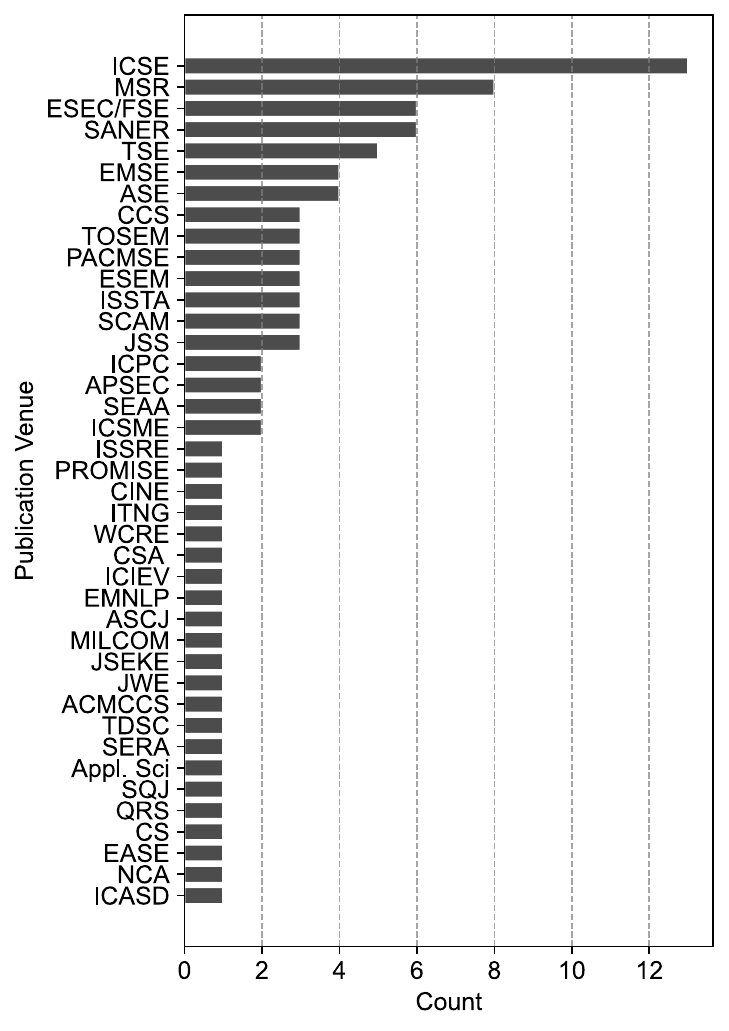}
  \caption{Publication venues.}
  \label{fig:publicationvenues}
\end{figure}
\subsection{\RqOne}
\label{RqOneResult}
\textit{Year of publication.} Figure~\ref{fig:publicationtrend} shows the analyzed primary sources span over two decades, from 2004 to May 2025. To analyze publication trend, we used a two-year span instead of a single year because the publication process commonly takes multiple years, making one-year counts an unreliable measure of output. Thus, our trend analysis indicates that the number of publications increases over time, with
notable peaks in 2022-2023 and a decline in 2024–May 2025. The latter decrease is common in secondary studies and is most likely due to incomplete indexing of recent publications by the bibliographic database.

Overall, these observations suggest a growing and sustained research interest of exploiting commit messages for corrective software maintenance: the topic remains active and evolving.

\textit{Venue.} Publication venues are quite varied, reflecting the nature of research on corrective maintenance and the shared interests of various communities.
As shown in Figure~\ref{fig:publicationvenuestypes}, the research community published their works more through conferences (71) than journals (26). While research papers are published across a wide range of venues, most works are published in leading software engineering international conferences like ICSE (13), MSR (8), ESEC/FSE (6), and SANER (6), as well as in the top journals like TSE (5), EMSE (4), TOSEM (3), and JSS (3) reflecting a particularly strong contribution from the empirical software engineering community as shown in Figure~\ref{fig:publicationvenues}.
However, other venues appear only once or twice in the dataset, showcasing the broad distribution of studies throughout the publication landscape. This suggests that using commit messages for corrective maintenance is highly relevant to software engineering domains and concentrated in top conferences and journals.

\subsection{\RqTwo}

\begin{figure}[t]
  \centering
  \includegraphics[width=1\columnwidth]{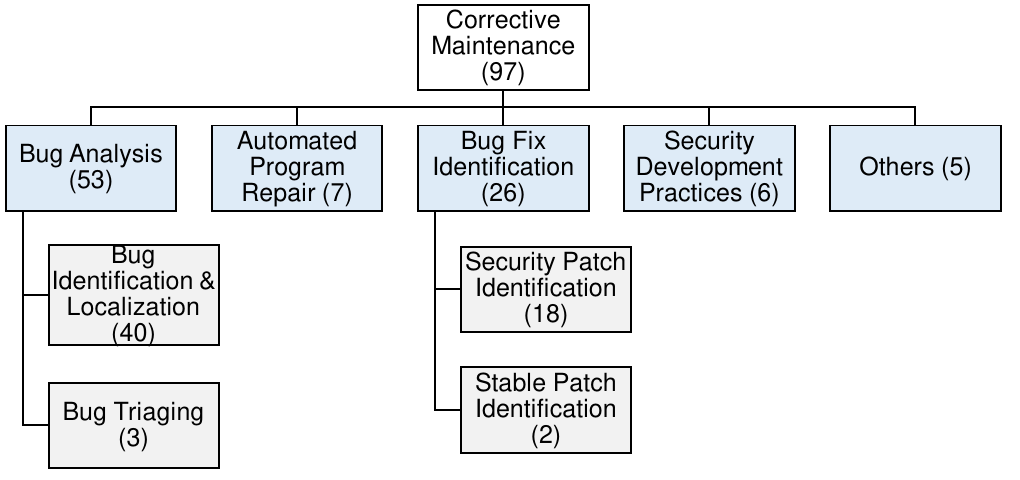}
  \caption{Taxonomy of research goals where the goal categories are clustered under four major themes (plus Others).}
  \label{fig:researchgoal}
\end{figure}
Since our focus is on the use of commit messages for corrective maintenance, it is essential to understand which goals researchers pursue with them. In this research question, we particularly review the specific goals of executing research on corrective maintenance. Our analysis led us to identify eight categories of corrective maintenance activities as shown in Table~\ref{tab:researchcategories}. 

\begin{table*}[t]
\centering
\caption{Research goal categories and their relevant papers from primary source.}
\label{tab:researchcategories}
\begin{tabular}{@{}p{2.8cm} p{12cm} p{2.2cm}@{}}
\toprule
\textbf{Categories} & \textbf{Goals} & \textbf{List of Papers} \\
\midrule
Bug analysis & Analysis of compiler bug propagation, dormant bugs vs non-dormant bugs, atom of confusion, reverted commits, security vulnerabilities, root causes of program bug,  defect proneness, relationships between stability and bug-proneness of clones, correlation between commits social characteristics and bugginess, etc. & \citePrimary{pinheiro2023they, yan2019characterizing, walkinshaw201820, rahman2017relationships, zhang2018empirical, barnett2016relationship, zhong2025understanding, eyolfson2011time, chen2014empirical, jimenez2018enabling} \\\hline
Bug identification and localization & Just-in-time defect prediction and localization, vulnerable dependency prediction, impact of tangled code changes on defect prediction models. In addition, this category also include optimization related topics such as investigation of “unfair, imbalanced” datasets effect on prediction performance, impact of misclassifications of bug reports in prediction, quality of attributes in predicting defect, heuristic to improve fault prediction,  improve tracing of defects through new SZZ algorithm, evaluate commits that introduce bug, searching code that is semantically identical to given buggy code and assessing practitioner beliefs about software defect prediction, etc. & \citePrimary{jiang2025just, abu2024parameter, liu2023semantic, sun2023silent, guo2023study, bludau2022pr, choi2023just, zeng2021deep, zhang2019commit, hoang2019deepjit, eken2019predicting, caulo2019use, suzuki2017application, tasnim2018inferring, ratzinger2007quality, tang2025llm4szz, herzig2016impact, kim2008classifying, tan2015online, chen2023boosting, ni2022best, zhou2025bridging, herzig2013s, guo2023code, sliwerski2005changes, bhattacharya2012graph, muthukumaran2015mining, perl2015vccfinder, wattanakriengkrai2020predicting, shrikanth2020assessing, jiang2013personalized, li2012cbcd, zimmermann2007predicting, bao2022v, herzig2013predicting, zhou2022simple, bird2009fair, jimenez2019importance, shivaji2012reducing, youm2015bug} \\\hline
Bug triaging & Assessing bug report and recommending the most appropriate developer for bug resolution. & \citePrimary{alkhazi2020learning, arnob2025bug, hossen2014amalgamating} \\\hline
Automated program repair & Improving patch quality and correctness,  vulnerability comprehension in automated program repair and  thereby reducing manual efforts. & \citePrimary{ahmed2023better, tian2022change, bai2021jointly, gao2025teaching, tufano2019empirical, lutellier2020coconut, tian2017automatically} \\\hline
Bug fix identification & Identifying commits that fix bugs. In addition, this category also include optimization related topics such as improving classification of bug fix commits, analyze data quality that have influence on bug fixing process, relation of bug fix and commit size, etc.  & \citePrimary{amit2021corrective, zafar2019towards, tian2012identifying, marzban2011cohesion, bachmann2010process, murgia2010machine} \\\hline
Security patch identification & Security patch identification/localization, hidden security fix pattern identification, commits that fix bugs with security implications and alert users to silent fixes, characterize the nature of security fixes in comparison to other non-security bug fixes, etc. & \citePrimary{zhang2024dual, farhi2025patchview, nguyen2022hermes, zhou2023tmvdpatch, nguyen2022vulcurator, zuo2023commit, sun2023exploring, shen2023patchmatch, wu2022enhancing, hoang2019patchnet, wang2022vcmatch, wang2021patchrnn, zhou2017automated, zhou2021spi, tan2021locating, li2017large, sabetta2018practical, sawadogo2022sspcatcher} \\\hline
Stable patch identification & Identifying stable patches that can be used for long term. & \citePrimary{liu2024patchbert, liu2025patchscope} \\\hline
Security development practice [Human Aspect] & Understanding why developers introduce vulnerabilities, determining where developers document security concerns, informativeness of security commit message, defining systematic scheme for identifying security vocabularies, creation of a convention for security commit messages that structure and integrate information about vulnerabilities. & \citePrimary{mock2024developers, reis2023security, fulton2022understanding, reis2022secom, antal2020exploring, morrison2018identifying} \\\hline
Others & Goals that do not fit above categories such as understanding relation between developer sentiment and software bugs, dependency management practice in response to bug, etc. & \citePrimary{cogo2019empirical, huq2020developer, hoang2020cc2vec, chen2020machine, huq2019understanding} \\
\bottomrule
\end{tabular}
\end{table*}

Figure~\ref{fig:researchgoal} illustrates the frequency of papers related to each corrective software maintenance category.
We observed that the goals motivating to undertake research activities can be clustered under four major themes: bug analysis (53), automated program repair (7), bug fix identification (26), and security development practices (6). Among these, the bug analysis category comprises two subcategories---bug identification and localization, and bug triaging---while the bug fix identification category includes subcategories security patch identification and stable patch identification. The majority of the papers fall under bug identification and localization (40), revealing a strong emphasis on enhancing software reliability through early stage bug identification during software development. The next dominant subcategory is security patch identification (18), which aims to identify hidden security fixes and alert users about their nature.
However, other categories such as automated program repair (7), security development practices (6), bug triaging (3), and stable patch identification (2) received relatively less attention compared to the dominant categories. 

\subsection{\RqThree}
\begin{figure*}[t]
  \centering
  \includegraphics[ width=0.72\linewidth]{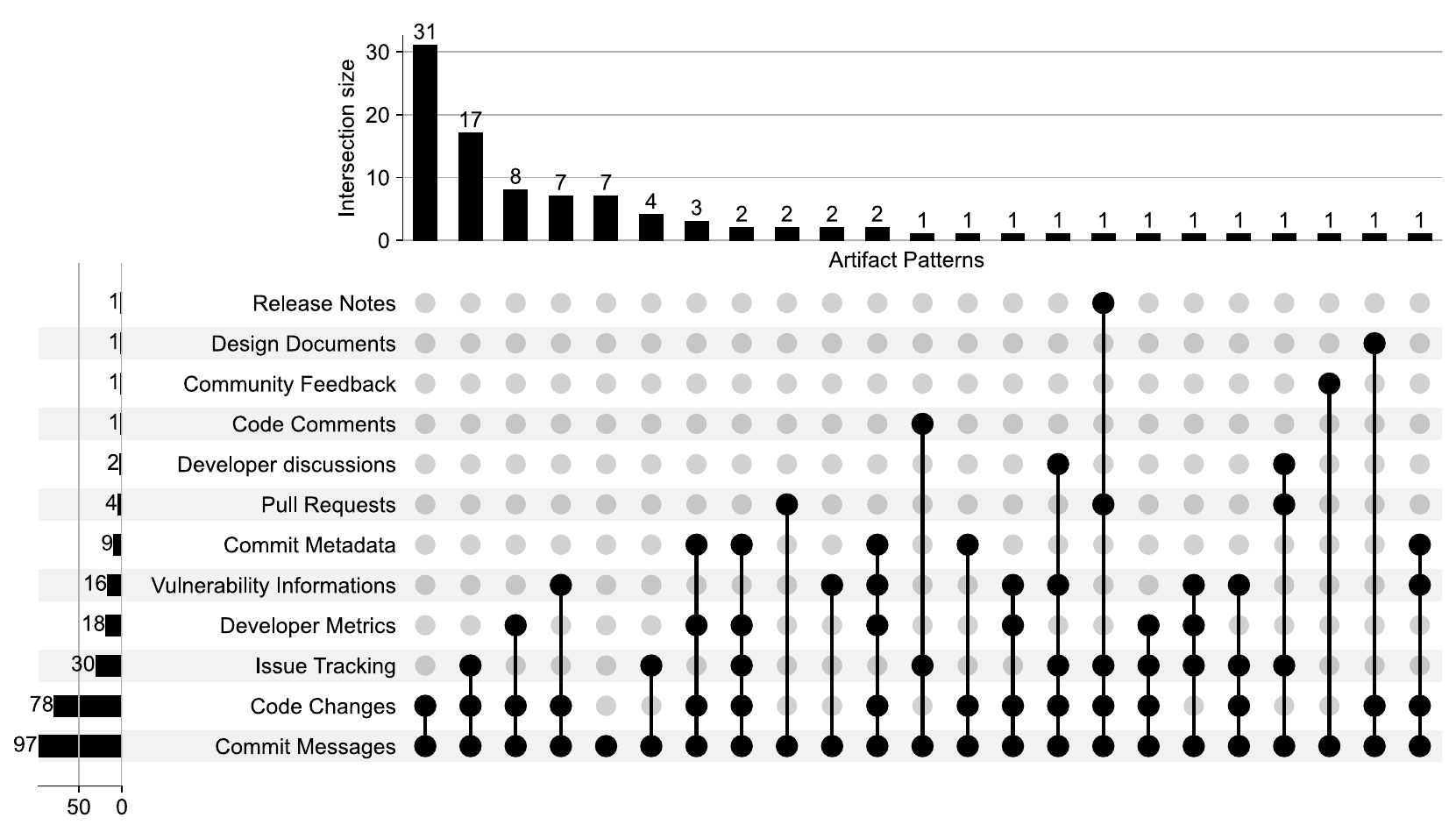}
  \caption{Frequency and co-occurrence of software artifacts used to support corrective software maintenance activities.}
  \label{fig:artifacttype}
\end{figure*}

Our analysis of what software artifacts were used in the reviewed studies, together with commit messages, identified 11 types of artifacts:
\begin{itemize}
    \item \textit{Code changes: } includes diff/patch/code changes of a commit, lines of code (LOC)/churn per commit within diff, etc.
    \item \textit{Issue tracking: } includes information extracted from issue/bug tracking systems, such as GitHub issues, Jira tickets, Bugzilla reports, etc.
    \item \textit{Developer metrics: } includes number of commits by author in past, number of files committed by commit author in the past, co-changed files, number of switches in projects by the author, ownership of files, work experience, and membership status such as owner, subscriber, watcher, stranger, etc. 
    \item \textit{Vulnerability information: } includes information about security vulnerabilities, in general from security advisories such as NVD, SNYK, GitHub, GitLab, reserved CVEs, etc.
    \item \textit{Pull requests: } includes data and metadata from pull and merge requests such as description, discussions, status, code changes (even when not merged in the target repository at the end), etc.
    \item \textit{Developer discussions: } includes content from question and answer web sites like stack overflow, mailing list, etc.
    \item \textit{Community feedback: } includes feedback from specific community through surveys, and in some cases also in-person workshops.
    \item \textit{Release notes: } includes information extracted from the release notes of specific software versions or packages.
    \item \textit{Code comments: } includes code comments extracted from source code.
    \item \textit{Design documents: } includes team members document on design decision. 
    \item \textit{Commit metadata: } includes commit information other than messages and diffs (already covered in other categories above), such as timestamps, commit hash identifiers, authors, etc.
\end{itemize}
As shown in Figure~\ref{fig:artifacttype}, code changes (78) artifacts emerged as the most used, followed by issue tracking information (30), developer metrics (18), and vulnerability information (16).
This dominance suggests that code changes play a central role in shaping the observed research patterns in artifact usage while executing corrective maintenance related studies---together with commit messages, of course.
Artifact issue tracking contents, while less prevalent than code changes, frequently co-occur with code changes, indicating a potential interdependence between the two (and commit messages). Artifact developer metrics and vulnerability information from security advisories, though less dominant overall, appear to complement the roles of both commit messages and code changes in specific contexts. The most frequently observed combination pattern within the dataset is $\{$commit message, code changes$\}$ followed by $\{$commit message, code changes, issue tracking$\}$, which recurred across most patterns as shown in Figure~\ref{fig:artifacttype}. This recurring pattern highlights the significance of their joint application in achieving desired research goals in corrective maintenance tasks. Conversely, the isolated use of commit messages appeared less common, implying a lower effectiveness in supporting corrective maintenance activities by themselves. Artifacts such as code comments and design documents were observed less frequently and often appeared in specialized scenarios, indicating their evolving relevance within the broader framework.

\begin{figure}[t]
  \centering
  \includegraphics[ width=.83\columnwidth]{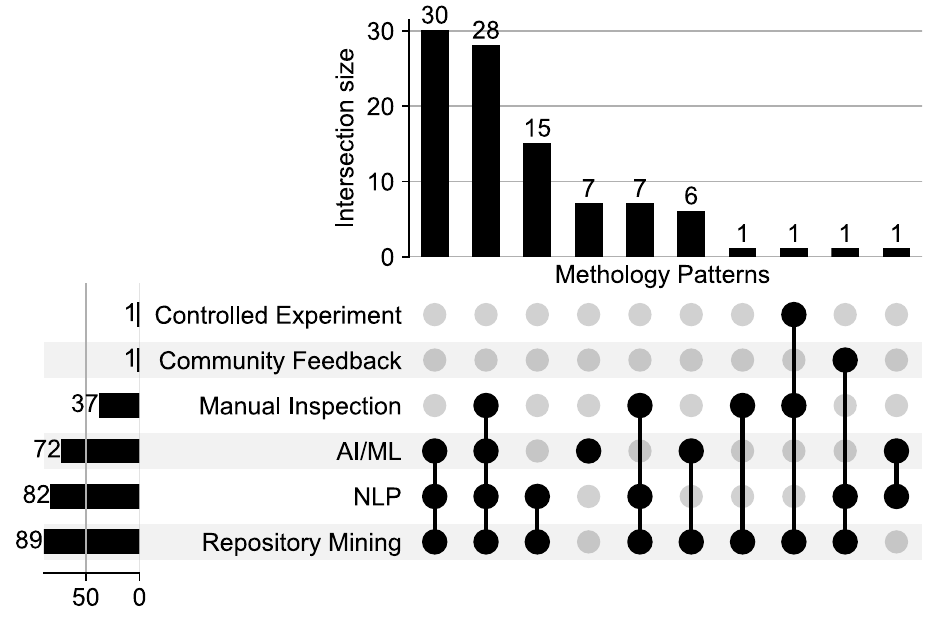}
  \caption{Methodology types and their usage patterns.}
  \label{fig:methodologyttype}
\end{figure}
%\balance
\subsection{\RqFour}
As shown in Figure~\ref{fig:methodologyttype}, our analysis revealed six primary methodology types: repository mining, NLP, AI/ML, manual inspection, community feedback, and controlled experiment. Among these, repository mining (89) emerged as the most dominant, followed by NLP (82) and AI/ML (72) techniques. The prevalence of repository mining indicates its broad applicability and perceived effectiveness across corrective maintenance related research settings. NLP and AI/ML techniques often appeared in conjunction with repository mining, suggesting their combined use provides complementary strengths and enhances the overall rigor of research outcomes. Manual inspection, while less frequent, was typically employed in studies requiring deeper analytical or multi-dimensional investigation on data. The combination $\{$ Repository mining, NLP, AI/ML$\}$ is the most common methodological pattern, reinforcing the view that integrating these approaches yields more comprehensive insights. In contrast, community feedback and controlled experiments were applied less frequently, reflecting their limited usage for corrective maintenance research contexts. Overall, the observed patterns collectively highlight a structured hierarchy, where repository mining serves as the methodological foundation, often supported by NLP and AI/ML, occasionally complemented by manual inspection.

\subsection{\RqFive}
Our analysis revealed three primary stakeholder types explicitly mentioned across the reviewed primary studies: developer, maintainer and researcher. Below, we define each stakeholder similar to previous work~\cite{hossen2014amalgamating} as follows:
\begin{itemize}
    \item \textit{Developer:} is mainly responsible for building software through development activities such as writing and modifying source code, fixing bug, testing components, etc. In addition, handling security responsibilities such as implementing security controls are integrated part of development role~\citeReferences{wang2021characterizing}. Therefore, security engineers and developers are considered as the same person and used interchangeably.
    \item \textit{Maintainer: } is mainly responsible for the governance of software project such as crucial decision making, reviewing contributions and merging, bug triaging, ensure code quality and consistency, etc.
    \item \textit{Researcher: } is someone who investigates software-related phenomena using scientific methods to produce new knowledge, tools, or validated theories.
\end{itemize}
\begin{figure}[t]
  \centering
  \includegraphics[width=0.75\columnwidth]{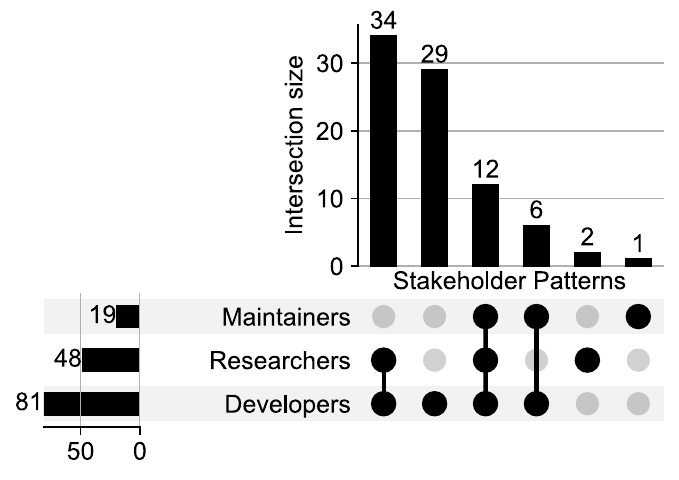}
  \caption{Stakeholder types and their mention patterns.}
  \label{fig:stakeholderstype}
\end{figure}
As shown in Figure~\ref{fig:stakeholderstype}, developers (81) appeared most frequently, indicating their central role in corrective software maintenance.
Researchers (48) are next, often considered alongside developers, suggesting the importance of applied research and collaboration among researchers and practitioners.
Maintainers (19) were considered last, typically in studies emphasizing stability and security-focused perspectives, indicating their responsibility for maintaining system security and compliance.
Overall, the findings indicate clear dominance of developers, underscoring how attention is often given to stakeholders who play the most critical role in shaping decisions and improving corrective maintenance practices. 

\subsection{\RqSix}
The key findings of the previous studies about how commit messages influence corrective maintenance are mapped into four major themes. These are explained as follows.

\textit{In bug analysis context,} commit messages play a vital role in tasks such as bug analysis, prediction, and triaging. Previous studies reported that commit messages convey important semantic information that helps machine learning models to better understand code changes and improve Just-in-time defect prediction accuracy~\citePrimary{jiang2025just, abu2024parameter}. Integrating commit messages with code change enhances bug detection performance, whereas the absence of either commit messages or code changes significantly reduces model accuracy. 
In addition, detailed and high-quality messages allow models to identify bugs more efficiently, while short or noisy ones negatively impact prediction outcomes~\citePrimary{ liu2023semantic, sun2023silent, guo2023study, choi2023just, zeng2021deep, zhang2019commit, barnett2016relationship, suzuki2017application}. Moreover, studies on bug analysis also reported that commit messages often lack sufficient details~\citePrimary{pinheiro2023they, bludau2022pr}. In bug triaging, the linguistic features of commit messages align closely with bug reports, helping to assign issues to the right developers and improving triaging accuracy~\citePrimary{alkhazi2020learning, arnob2025bug, hossen2014amalgamating}.
Overall, informative and structured commit messages enhance both prediction accuracy and the interpretability of automated tools in managing software bugs.

\textit{In automated program repair context, } commit messages contribute valuable insights into the intent and context behind code changes. Well-structured commit messages help large language models (LLMs) learn to produce better patches, effectively linking natural language description with code semantics~\citePrimary{ahmed2023better, tian2022change}. Studies on automated program repair also reported that commit message quality correlates with correct patch identification and recurring linguistic patterns within commit messages provide features that improve automated analysis and repair~\citePrimary{gao2025teaching}. 
Overall, high quality commit messages not only document developer intent but also act as a rich data source to strengthen the precision and interpretability of automated program repairing techniques.

\textit{In bug fix identification context,} commit messages are essential, but many of them lack informativeness, limiting the automated detection accuracy~\citePrimary{zhang2024dual, amit2021corrective, zafar2019towards}. Commit messages that are clear and descriptive, with fix-related keywords or issue references, enable tools to distinguish bug-fix commits from regular updates and support efficient maintenance tracking~\citePrimary{bachmann2010process, murgia2010machine}. For security patch identification, well-crafted messages containing security-related keywords such as CVE ids combined with code diffs are especially useful in recognizing vulnerability-fixing commits and significantly improve model accuracy~\citePrimary{nguyen2022hermes, zhou2023tmvdpatch, nguyen2022vulcurator, zuo2023commit, sun2023exploring}.
Overall, precise and detailed commit messages with specific vocabulary enhance the automation and reliability of identifying bug fix related commits across various contexts.

\textit{Security development practice related studies,} reported that commit messages often fail to provide sufficient security-related details, limiting their utility for vulnerability assessment~\citePrimary{mock2024developers, reis2023security}. Most commit messages lack explicit references to security issues, relying on implicit domain vocabulary instead~\citePrimary{morrison2018identifying}.  Well-structured and informative messages that explicitly describe security implications can facilitate automated vulnerability detection and human understanding~\citePrimary{reis2022secom, fulton2022understanding}. Beyond security, other studies show that commit messages capture developer sentiment, and rationale for code changes, which can influence software quality and team collaboration~\citePrimary{huq2020developer, huq2019understanding}.
Overall, improving the linguistic richness, structure, and developers apparent sentiment (as detectable by sentiment analysis) in commit messages can strengthen both technical and human aspects of software development.

\section{Recommendations}
\label{sec:implication}

Building on the results above, we present key recommendations for stakeholders (such as developers, maintainers, and researchers) to improve commit message quality and, eventually, corrective software maintenance.

\textit{Developers} are recommended to follow strict rules and lexicons when writing commit messages, ensuring semantic alignment with code changes~\citePrimary{guo2023study, zeng2021deep, caulo2019use}. In addition, their written messages should preserve essential natural-language patterns and technical terminologies, as they serve as a bridge between code-level changes and natural language understanding~\citePrimary{gao2025teaching}.   For bug fix identification purposes, developers are recommended to include explicit rationale and contextual explanations in commit messages to make maintenance more efficient~\citePrimary{bachmann2010process}. Besides, keeping commit messages concise with separate commit-info, commit-subject, and commit-body can improve automated understanding and accuracy of patch identification ~\cite{zhou2021spi}. For better security patch identification, developers should avoid hiding the nature of security vulnerability in commit message~\citePrimary{nguyen2022hermes}. Besides, messages should clearly describe the type of security issue addressed by including references (like CVE or bug IDs) to facilitate easier linking of fixes~\citePrimary{zuo2023commit, wang2022vcmatch, tan2021locating}. For developers, these recommendations can help understand necessary technical knowledge to write better commit messages and supporting later corrective software maintenance.

\textit{Maintainers} should be conscious that the quality of commit messages will directly impact later corrective maintenance tasks.
When starting a new project, they should provide explicit guidelines and possible explanations (e.g., by providing good examples and better documentation) to write commit messages, ensuring semantic alignment with code changes.
For instance, a univocal language can be used to identify the type of operation committed by developers, such as bug fix, refactoring, etc.~\citePrimary{guo2023study, zeng2021deep, caulo2019use}. Recent conventions about how to write commits, like Conventional Commits~\citeReferences{ccs2025}, can help in this respect. In addition, maintainers should be proactive in reviewing commits made by newbies carefully and ensure that code changes are accompanied by well-structured and detailed messages~\citePrimary{suzuki2017application}. For maintainers, such understandings help support cost-effective maintenance and evolution of their software projects.

\textit{Researchers } should standardize templates, official guidelines, and best practices for writing commit messages, as these make it easier for automated tools and human analysts to detect bugs~\citePrimary{reis2022secom}. They should encourage developers to adopt best development practice (such as security), through highlighting how informative commit messages with structured and consistent formatting can facilitate bug assessment~\citePrimary{reis2023security}. For instance, a study by~\citePrimary{morrison2018identifying} suggested that combining standard security keywords (e.g., “encryption,” “authentication”) with project-specific vocabulary further improves precision in identifying security-relevant changes. In building automated classification, researchers are advised to be careful during pre-processing of commit message in order to avoid overlooking acronyms such as VRSN (version), RLS (release), etc~\citePrimary{murgia2010machine}.
To further improve bug classification performance and reduce the false positive rate, researchers should focus on enriching commit messages with supplementary information, such as test data, issue tracking contents, etc~\citePrimary{herzig2013s}. Moreover, researchers should investigate possible ways to reduce manual verification effort, as previous studies further suggest that even when automated classifiers identify security commits, manual expert verification remains essential for ensuring precision~\citePrimary{sabetta2018practical}.  Beyond these, other studies recommend that researchers should also explore the sentiment and emotional cues in commit messages to understand developer performance and team collaboration dynamics~\citePrimary{huq2019understanding}.

\section{Threats to validity}
\label{sec:threats}
Like all secondary studies, threats to validity can be raised by the research methodology applied by this systematic mapping study. We have mitigated
such threats by following the systematic review protocol proposed by~\citetReferences{wohlin2012experimentation}.

Threats to internal validity may be introduced by having inappropriate classification and interpretation of the studies. We have limited this by conducting three iterations of paper reviews by the two authors collaboratively. In addition, unclear data items are revisited and resolved through discussions and further
research during the search.

Threats to construct validity may be introduced by inadequate capture of relevant literature for the target study.
Our initial search string does not include \emph{specific} terms related to corrective maintenance (e.g., ``defect'', ``repair'', ``bug'', etc.), but rather retrieves studies broadly related to commit messages.
This choice was due to the difficulty of being comprehensive with specific terms; short of which the initial seed of papers might have been imbalanced, favoring some concepts over others.
In addition, we do not consider the popularity of conference and journal venues as a selection criterion, which may result in the inclusion of a few studies from less widely recognized venues; on the other hand, this choice protects against venue-related biases.

Threats to external validity may restrict the generalizability of the revealed statistics and the justification of future research opportunities.
Since our study was focused on peer-reviewed publications, some relevant industrial or unpublished work (e.g., gray literature) may have been missed.
Furthermore, despite leveraging multiple digital libraries and following a rigorous protocol to craft search queries, we might still have missed relevant studies.
The inclusion of a backward snowballing step mitigates this risk to some extent, but complete coverage is never guaranteed in any systematic review of the literature.
This, together with evidence from best practices in corrective software maintenance, has provided rich sources for us to discuss the results and the future research opportunities identified.

\section{Conclusion}
\label{sec:conclusion}
In this paper, we conducted a systematic mapping study on the usage of commit messages in corrective software maintenance
accompanied by a meta-analysis to answer our research questions. After a comprehensive search that follows a systematic series of steps and assessing the quality of the studies, 97 papers were identified.

Based on the data extracted from these selected papers, we derived a comprehensive synthesis on the state-of-the-art of the use of commit messages for corrective software maintenance tasks.
We observed a consistent and growing interest in the usage of commit messages to resolve issues of corrective maintenance over the last two decades, with a marked increase in 2022-2023.
Although bug analysis is the dominant goal, relatively few studies address other themes such as automated program repair, security development practices, etc. We have observed $\{$commit messages, code changes$\}$ to be the most dominant artifact co-usage pattern. In terms of methodologies: repository mining, followed by NLP, and AI/ML techniques are the most popular ones. 
Our analysis demonstrates that commit messages play a significant role in corrective maintenance activities since they carry crucial information helpful for stakeholders (i.e., developer, maintainer, and researcher) to understand and improve the code base through the software evolution process. However, commit messages often lack necessary information and are not always sufficient to represent the intent of code changes.

Among the gaps, we have observed in the literature, three stand out: (1) the need to close the gap between stated goals of corrective maintenance and stakeholders' needs; (2) a thorough study to investigate the trade-offs between benefits and drawbacks of following official guidelines in writing commit messages, to identify what an ideal project VCS history---including commit messages---would look like;  (3) interviewing or surveying stakeholders to further allow for triangulation of our findings to understand the impact of commit message quality on corrective maintenance tasks.

\section*{Data availability}
A complete reproducibility package for the work described in this paper is publicly archived and available from Zenodo~\citeReferences{Replication2026}.

\begin{acks}
This work was supported by France Agence Nationale de la Recherche (ANR), program France 2030, reference ANR-22-PTCC-0001.
The authors would like to thank Théo Zimmermann for insightful discussions and comments on an earlier version of this paper.
\end{acks}

\bibliographystyleReferences{ACM-Reference-Format}
\bibliographyReferences{reference}   

\bibliographystylePrimary{ACM-Reference-Format}

\bibliographyPrimary{primary}   

\end{document}